\documentclass[letter]{aa}

\usepackage{graphicx,txfonts,natbib}
\bibpunct{(}{)}{;}{a}{}{,}

\begin{document}

\title{An Intriguing Solar Microflare Observed \\ with RHESSI, Hinode and
TRACE}

\author{I. G. Hannah\inst{1} \and S. Krucker\inst{1} \and H. S.
Hudson\inst{1} \and S. Christe\inst{1,2} \and R. P. Lin\inst{1,2}}

\offprints{I. G. Hannah \email{hannah@ssl.berkeley.edu}}

\institute{Space Sciences Laboratory, University of California at Berkeley,
Berkeley, CA, 94720-7450, USA \and Physics Department, University of
California at Berkeley,  Berkeley, CA, 94720-7450, USA}

\date{Received 8 Nov 2007; Accepted 3 Dec 2007}

\abstract{}{Investigate particle acceleration and heating in a solar
microflare.}{In a microflare with non-thermal emission to remarkably high
energies ($>50$ keV), we investigate the hard X-rays with RHESSI imaging
and spectroscopy and the resulting thermal emission seen in soft X-rays
with Hinode/XRT and in EUV with TRACE.}{The non-thermal footpoints
observed with RHESSI spatially and temporally match bright footpoint
emission in soft X-rays and EUV. There is the possibility that the
non-thermal spectrum extends down to 4 keV. The hard X-ray burst clearly
does not follow the expected Neupert effect, with the time integrated hard
X-rays not matching the soft X-ray time profile. So although this is a simple
microflare with good X-ray observation coverage it  does not fit the
standard flare model.}{}

\keywords{Sun:Corona - Sun:Flares - Sun: X-rays, gamma rays}

\titlerunning{RHESSI, Hinode and TRACE Microflare}

\authorrunning{Hannah et al.}

\maketitle

\section{Introduction}

A Solar flare is the observed multi-wavelength atmospheric response to a
rapid transient release of magnetic energy from coronal fields. This
process is observed over many scales with small flares, namely
\emph{microflares}, exhibiting energies expressed as millionths of that in
the largest flares. The general scenario of a flare is that accelerated
particles leave a coronal region where they have gained energy liberated
from the coronal magnetic fields, facilitated by magnetic reconnection.
These particles stream along the magnetic fields and are seen in hard
X-rays via thick target collisional bremsstrahlung at the loop footpoints
where the local density is large enough to stop the particles. As well as
creating hard X-rays, these accelerated electrons heat the local plasma
which evaporates into coronal loops, seen initially in soft X-rays  and
eventually in EUV as the material cools. The Neupert effect, the time
integrated hard (non-thermal) X-ray time profile matching the soft
(thermal) X-ray time profile, is often taken as a signature of this behavior
\citep{neupert1968,hudson1972,dennis1993,veronig2005}.

\begin{figure}[h]\centering
    \includegraphics[scale=0.38]{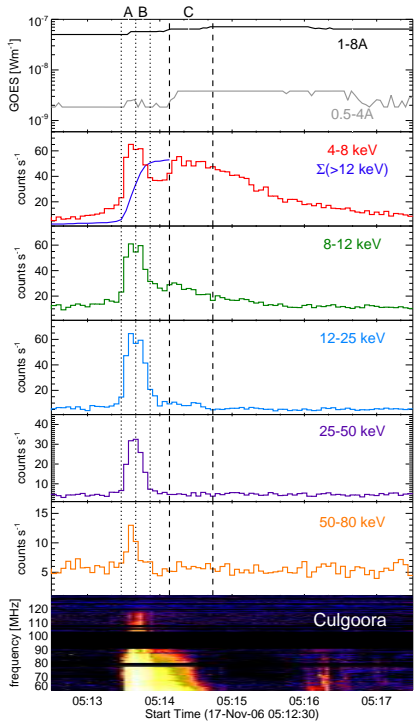}
\caption{\label{fig:rhessi_lc}
X-ray and radio time profiles for the 05:13 November 17, 2006 microflare,
showing the GOES soft X-ray emission and RHESSI observations in five
energy ranges. The 4-8 keV lightcurve is overplotted with the scaled
integrated non-thermal emission $> 12$ keV. The vertical dotted and
dashed lines indicate three time periods for which the RHESSI images and
spectra have been analyzed: two 12 second impulsive intervals (A and B)
and a 36 second thermal period (C). The spectrogram in the bottom panel
is Culgoora radio data, showing a strong Type III burst.} \end{figure}

In this paper we present {an} analysis into this scenario using a GOES
A-Class microflare where the Neupert effect is not observed, despite good
hard and soft X-ray coverage: the event was observed in soft and hard
X-rays with  RHESSI \citep{lin2002}, the X-ray telescope XRT \citep{xrt} on
Hinode \citep{hinode} and in EUV with TRACE \citep{trace}. The microflare
analyzed in this paper starts at 05:13:28 November 17, 2006 and shows a
remarkably hard/flat spectrum to high energies ($>50$ keV). This is
unusual as microflares observed with RHESSI typically have soft/steep
spectra observable only up to about 25 keV \citep{mfpart2}. This
microflare is from active region AR10923 which produced several hard
microflares, with a partially occulted one suggestive of thin target
emission \citep{krucker2007}. This active region also produced many Type
III radio bursts, a further signature of accelerated electrons
\citep{bastian1998}.

\section{Observation \& Data Analysis}

There is uninterrupted RHESSI coverage over the whole of this microflare.
However by November 2006, radiation damage to RHESSI's detectors had
resulted in an increased background and degraded spectral resolution.
This damage affects each of the 9 detectors differently and so we have
tried to use those least affected in the analysis: detectors 1,4,6 for
spectral analysis and detectors 3,4,6,8,9 for imaging. The damage still
influences the results, particularly the spectral analysis below 10 keV,
discussed further in \S\ref{sec:spectra}. During the microflare there are
XRT images every 20 seconds, alternating between the C-Poly and Ti-Poly
filters, with exposure times between 0.1 and 0.5 seconds. The Ti-Poly were
taken using a lower resolution (2") compared to the C-Poly images (1").
These filters have similar temperature responses, mostly sensitive to $<
10$MK plasma. TRACE images are also available for this event, with the
195\AA~ and 284\AA~ filters. The 195\AA~ images are every $\approx$ 45
seconds over this event, whereas only two 284\AA~images are available,
one pre-flare, the other during the flare. In addition there was also a bright
Type III radio burst associated with the flare impulsive peak observed by
various radio instruments, such as WIND/WAVES \citep{windwaves},
STEREO/WAVES \citep{stereowaves} and Culgoora. A subsequent increase
in interplanetary energetic particles was also observed about 30 minutes
after the microflare with WIND/3DP \citep{wind3dp}.

\begin{figure*}\centering
    \includegraphics[scale=0.7]{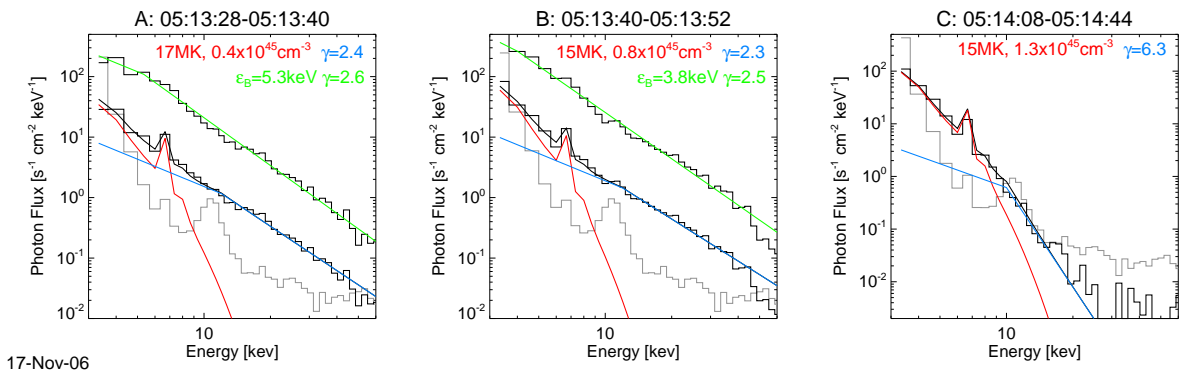}
\caption{RHESSI hard X-ray spectra for the three time intervals shown in
Figure \ref{fig:rhessi_lc}. The black and grey binned lines indicate the
background subtracted data and the background emission. The red, blue
and black lines show the thermal, broken power-law and total (thermal
plus broken power-law) model. The top data line fitted with a green line in
the first two spectra is the data fitted with just a broken power-law model,
multiplied by 10 to make it distinguishable from the other spectrum. The
photon spectra are different for the two fits in the same time interval
because they are found from the same count spectra convolved with
RHESSI's detector response and the different model fits.\label{fig:hsispec}}
\end{figure*}

\subsection{Time Profiles \& Spectra}\label{sec:spectra}

The X-ray and radio time profiles for the microflare are shown in Figure
\ref{fig:rhessi_lc}.  The GOES soft X-ray lightcurve peaks at GOES A7 level
in the 1-8\AA~channel, or A2 level taking the pre-flare background into
account. At the lowest RHESSI energies (4-8 keV), there are 2 distinctive
peaks  of emission: a short impulsive one from 05:13:28 to 05:13:52 and a
second from 05:14:08 until nearly 05:17:00. The higher energy ranges
($>12$ keV) only show emission during the initial impulsive phase. This
impulsive initial emission extends to remarkably high energies ($>50$ keV)
and is closely associated with a Type III radio burst, shown in the bottom
panel of Figure \ref{fig:rhessi_lc}. These factors clearly indicate that above
12 keV the emission is mostly non-thermal. Normally the lowest energies
RHESSI observes are predominantly thermal, but the time profile of the 4-8
keV emission does not demonstrate the Neupert effect: the time
integrated non-thermal emission $>12$ keV, shown in Figure
\ref{fig:rhessi_lc}, does not match the observed peak in 4-8 keV. This
implies that either the 4-8 keV emission during the impulsive peak is
non-thermal or some other process is occurring other than simple
evaporation and heating. Although in the vast majority of flares the low
energy RHESSI observes is thermal in origin, non-thermal emission
dominating at these energies has been suggested in a previous event
\citep{sui2006}.

\begin{figure*} \centering
    \includegraphics[scale=0.7]{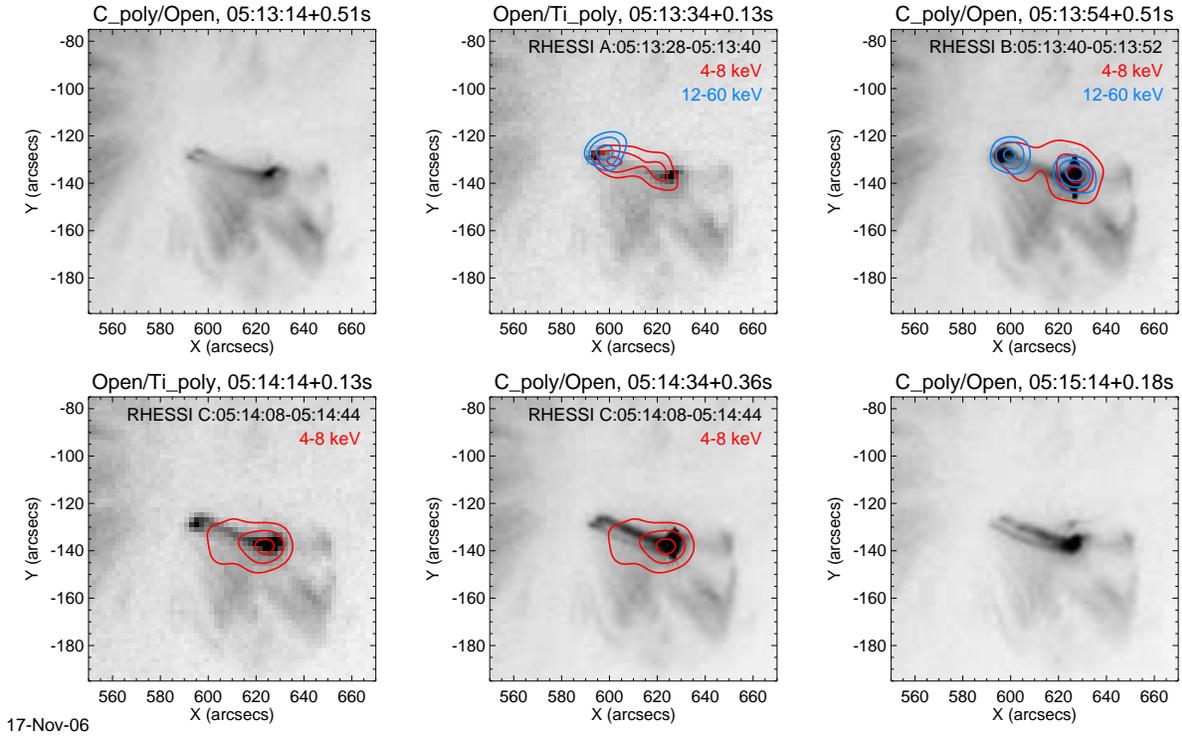}
\caption{\label{fig:xrtmem} X-ray imaging at six different times during the
05:13 November 17, 2006 microflare: (top left) pre-flare (top middle) first
12 second impulsive period (top right) second 12 second impulsive period
(bottom left \& middle) 36 second of thermal period and (bottom right)
post-flare. The background images are XRT with either the C-Poly or
Ti-Poly filters, the latter using a lower resolution. The number after the
time in the title of each image is the exposure time of each XRT image.
The overplotted contours (50, 70, 90\%) are the RHESSI images
reconstructed using the CLEAN algorithm with detectors 3,4,6,8 and 9. The
red contours represent the 4-8 keV emission (predominantly thermal) and
the blue contours are the 12-60 keV emission (mostly non-thermal). The
pointing of all the XRT images have been correct by x+5" and y+8". This
was found by matching the RHESSI hard X-ray footpoints with XRT bright
emission in the trop right image.} \end{figure*}

To investigate this in detail, the RHESSI {spectra} during three time
intervals have been analyzed: A 05:13:28 to 05:13:40, the rise phase of the
impulsive period, B 05:13:40 to 05:13:52 the decay phase of the impulsive
period, and C 05:14:08 to 05:14:44 the second peak. All the spectra during
these time intervals have been fitted with a thermal plus broken
power-law (to represent the non-thermal emission) model,  shown in
Figure \ref{fig:hsispec}. Additionally, the first two spectra, A,B during the
impulsive peak, have also been fitted by just a broken power-law
spectrum, indicated by the upper green line fit and multiplied by a factor of
10 to separate it from the other model fit shown. The resulting photon
spectra appear different for the two model fits as the conversion from
observed counts in the detectors to photons depends on the fitted model
convolved with RHESSI's non-diagonal detector response matrix. This
means that in the photon spectrum fitted with a thermal component the
thermal line features are over-emphasized. Normally the existence of a
thermal component is trivial to determine as the thermal line features are
directly detectable in the counts spectrum, indicating plasma $\geq 8$MK
\citep{phillips2004}, but due to the RHESSI's radiation-damaged detectors
these are not apparent. This damage also skews the resulting thermal fit,
producing slightly higher temperatures and lower emission measures than
expected \citep{mfpart2}. This resulting uncertainty causes an ambiguity
in the transition from where the the thermal dominates over the
non-thermal component and so the broken power-law is fitted using a
fixed break energy of 12 keV. The {spectra} are however, definitely very
flat/hard above this break, with $\gamma\approx 2.3 - 2.4$.

Assuming the power-law emission is thick-target the power in the
electrons can be estimated via \citet{brown1971}: the power in electrons
above 12 keV is $4\times10^{26}$ erg s$^{-1}$ and $3\times10^{26}$ erg
s$^{-1}$ over time intervals A and B. Over each of these 12 sec time
periods the total non-thermal energy is $8\times10^{27}$ erg. The purely
non-thermal broken power-law model produces slightly flatter spectrum,
though still very steep with $\gamma\approx 2.6 - 2.5$, with breaks down
to around 4 keV. The power in electrons above 4 keV in these {spectra} is
$2\times10^{27}$ erg s$^{-1}$ and $3\times10^{27}$erg s$^{-1}$ over
intervals A and B, estimating the total non-thermal energy as
$5\times10^{28}$ erg. The difference of a factor $\approx6$ in the total
non-thermal energy is a guide to the systematic errors in the estimate.

\subsection{X-ray \& EUV Imaging}

The time {progressions} of the RHESSI and XRT X-ray images for the
microflare are shown in Figure \ref{fig:xrtmem}. Pre-flare (05:13:14), XRT
observes 2 curved loops with their left footpoints a few arcseconds apart
and the right footpoints brighter and unresolvably close together. When
the flare begins the left footpoints in XRT {brighten} and RHESSI observes,
over time interval A, a single non-thermal 12-60 keV footpoint at this
location, as well as 4-8 keV emission across the XRT loop. As the impulsive
phase of the flare declines, time interval B,  both left and right footpoints
are bright in XRT, with the emission saturating at the right footpoints. The
RHESSI 4-8 keV emission again covers the XRT loop and now two 12-60
keV footpoints match the two bright XRT sources. During time interval C
only the 4-8 keV RHESSI emission is imageable, producing a single source
that matches the brightening righthand side of the loops seen in XRT. After
the flare emission ends in RHESSI, the two hot bright loops are still visible
in XRT, slightly further apart than pre-flare.

\begin{figure*} \centering
    \includegraphics[scale=0.7]{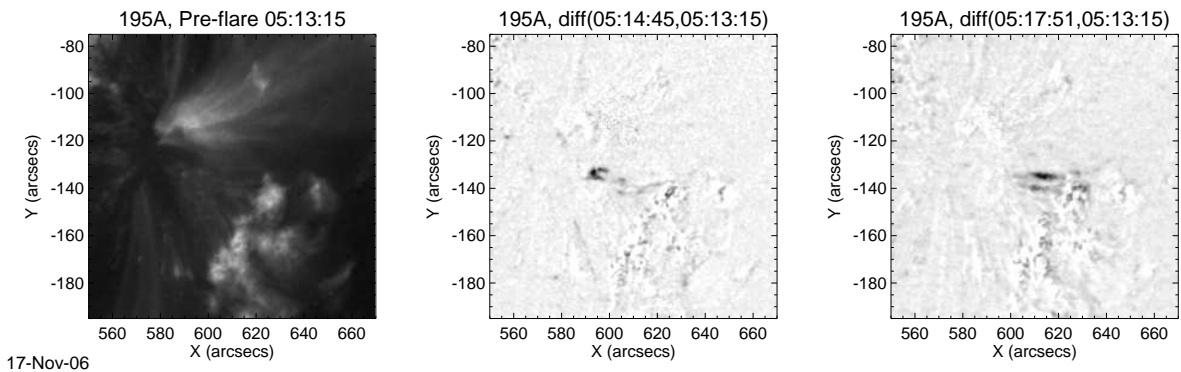}
\caption{\label{fig:trace195}TRACE 195\AA~ EUV images of the 05:13
November 17, 2006 microflare: (left) pre-flare image (middle) difference
image between {an} image during the RHESSI thermal period and the
pre-flare image (right) difference image between post- and the pre-flare
images. The normalization and order of the color scales varies between
the pre-flare image and the difference images.} \end{figure*}

The 195\AA~ TRACE images of this microflare are shown in Figure
\ref{fig:trace195}. Pre-flare no noticeable counterpart structures to those
seen in XRT are observed. At later stages of the flare and post-flare similar
behavior is observed to that in XRT but delayed. Namely the left footpoints
initially brighten with the later emission occurring from the righthand side
of the loops. This is consistent with the hot material initially seen by
RHESSI and XRT, eventually cooling below 2MK and being seen in EUV.
There is also a 284\AA~ TRACE image during the decline phase of the
RHESSI impulsive peak, time interval B. This image shows two bright
footpoints that closely match the bright 12-60 keV non-thermal footpoints
seen in RHESSI. This is indicative of the accelerated electrons, inferred by
the RHESSI observations, producing continuum observable in TRACE with
the 284\AA~ filter.

\begin{figure}\centering
    \includegraphics[scale=0.7]{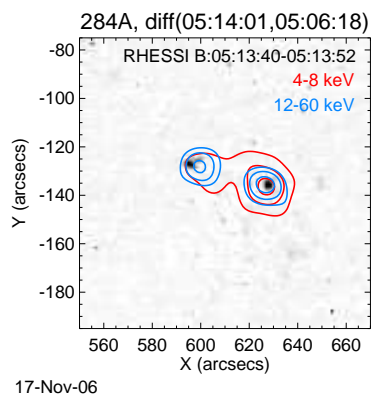}
\caption{\label{fig:trace284}
TRACE 284\AA~ EUV difference image during the 05:13 November 17,
2006 microflare. The difference image is between an image just after the
impulsive period and the pre-flare. Overplotted are the RHESSI thermal
(4-8 keV) and non-thermal (12-60 keV) contours from impulsive interval B.
The pointing of the TRACE image has been correct by x+3" and y+7". This
was found by matching the RHESSI hard X-ray footpoints with bright EUV
footpoints. } \end{figure}

\section{Discussion \& Conclusions}

The microflare presented here clearly shows electrons accelerated to
relatively high energies, resulting in heating that is observed in X-rays and
EUV. The lower energy RHESSI observations however do not demonstrate
the expected Neupert effect in their time profile, suggesting that the initial
4-8 keV emission may not be thermal. {Such hard flat spectra with
unexpectedly delayed thermal emission have been observed previously in
\emph{early impulsive flares} \citep{farnik1997,sui2007} but never in such
a small flare or with possible non-thermal emission to low energies ($<8$
keV). An alternative to the standard flare scenario for this microflare is
that the non-thermal emission observed in the initial impulsive peak
deposits its energy, resulting in chromospheric evaporation, but these
higher energy electrons penetrate deeper into the chromosphere, cooling
quickly in the dense lower chromosphere, resulting in a diminished upward
flow \citep{mcdonald1999}. This still leaves the low energy ($<8$ keV)
accelerated electron population, where the bulk of the non-thermal energy
input lies, which should produce detectable evaporation. This could be
responsible for the subsequent thermal emission, but the delay is still
intriguing and requires further study and modeling.}

One major consequence of the non-thermal emission extending down to
lower energies is an increased non-thermal energy input into the lower
atmosphere. If we assume that this non-thermal energy is deposited into
approximately the footpoint area in the XRT images, then the electron
fluxes can be estimated. During both time interval  A and B, the individual
XRT footpoint area is about $4"\times6"$ or $\lesssim 10^{17}$ cm$^{2}$.
With a thermal component at the lowest RHESSI energies then the fluxes
of electrons with $>12$ keV deposited is $4\times10^{9}$ erg cm$^{-2}$
s$^{-1}$, into a single footpoint during interval A, and $2\times10^{9}$ erg
cm$^{-2}$ s$^{-1}$, {in total} into the two footpoints during interval B.
Both of these values are indicative of \emph{gentle} chromospheric
evaporation \citep{fisher1985,abbett1999,brosius2004,milligan2006} since
the fluxes of flare-accelerated electrons are $\leq 10^{10}$ erg cm$^{-2}$
s$^{-1}$. If instead the RHESSI emission is purely non-thermal down to low
energies then the fluxes of electrons with $>4$ keV deposited is
$2\times10^{10}$ erg cm$^{-2}$ s$^{-1}$ and $1\times10^{10}$ erg
cm$^{-2}$ s$^{-1}$. These values are larger and could lie in the regime of
\emph{explosive} evaporation $\geq 3\times 10^{10}$ erg cm$^{-2}$
s$^{-1}$. During the second peak (interval C) the RHESSI time profile and
spectrum are more consistent with thermal emission. Using the RHESSI
emission measure from this time ($1.3\times10^{45}$ cm$^{-3}$) and
estimating the loop volume from the XRT image, ($32"\times3"$) giving
$10^{26}$ cm$^{-3}$, the {instantaneous} thermal energy (without losses)
over this time period can then be estimated as $2\times10^{27}$ erg
(calculation detailed in \citet{mfpart2}). Depending on whether thermal or
non-thermal energies dominate the RHESSI low-energy component during
the first impulsive peak, this thermal energy is either a 4 or 20 times
smaller than the non-thermal input. The difficulties in determining whether
the low energy emission is thermal or non-thermal is partially due to
RHESSI's degraded performance due to radiation damage. This should be
rectified with the November 2007 anneal and so 2008 onwards flare
studies should benefit from a reinvigorated RHESSI, as well as newer
instruments such as Hinode. The microflares of AR10923 however are still
very interesting, with further analysis and modeling required for these
microflares, especially into how electrons were accelerated to such high
energies in these small flares.

\begin{acknowledgements}

NASA supported this work under grant NAS5-98033 and NNG05GG16G.
Hinode is a Japanese mission developed and launched by ISAS/JAXA, with
NAOJ as domestic partner and NASA and STFC (UK) as international
partners. It is operated by these agencies in co-operation with ESA and
NSC (Norway).

\end{acknowledgements}


\end{document}